\documentclass[prd,superscriptaddress,twocolumn,nopreprintnumbers,
floatfix]{revtex4}
\usepackage{graphicx,url,amssymb,amsmath,longtable,rotating,color,units,
wasysym,subfigure,epsfig,multirow,amsbsy,soul,mathrsfs,amsmath,xcolor,mathtools}
\usepackage[plainpages=false, colorlinks=true, anchorcolor=blue, linkcolor=blue, citecolor=blue, bookmarks=false]{hyperref}

\newcommand{\GR}{GR}
\newcommand{\GRP}{$\text{GR}_+$}

\begin{document}

\title{Challenges testing the no-hair theorem \\
with current and planned gravitational-wave detectors}
\author{Eric Thrane}
\affiliation{Monash Centre for Astrophysics, School of Physics and Astronomy, Monash University, VIC 3800, Australia}
\affiliation{OzGrav: The ARC Centre of Excellence for Gravitational-wave Discovery}
\email{eric.thrane@monash.edu}
\author{Paul D. Lasky}
\email{paul.lasky@monash.edu}
\affiliation{Monash Centre for Astrophysics, School of Physics and Astronomy, Monash University, VIC 3800, Australia}
\affiliation{OzGrav: The ARC Centre of Excellence for Gravitational-wave Discovery}
\author{Yuri Levin}
\email{yuri.levin@monash.edu}
\affiliation{Department of Physics, Columbia University, New York, NY 10027, USA}
\affiliation{Center for Computational Astrophysics, Flatiron Institute, 162 5th Ave, New York, 10010, NY, USA}
\affiliation{Monash Centre for Astrophysics, School of Physics and Astronomy, Monash University, VIC 3800, Australia}

\begin{abstract}
General relativity's no-hair theorem states that isolated astrophysical black holes are described by only two numbers: mass and spin.
As a consequence, there are strict relationships between the frequency and damping time of the different modes of a perturbed Kerr black hole.
Testing the no-hair theorem has been a longstanding goal of gravitational-wave astronomy.
The recent detection of gravitational waves from black hole mergers would seem to make such tests imminent.
We investigate how constraints on black hole ringdown parameters scale with the loudness of the ringdown signal---subject to the constraint that the post-merger remnant must be allowed to settle into a perturbative, Kerr-like state.
In particular, we require that---for a given detector---the gravitational waveform predicted by numerical relativity is indistinguishable from an exponentially damped sine after time $t^\text{cut}$.
By requiring the post-merger remnant to settle into such a perturbative state, we find that confidence intervals for ringdown parameters do not necessarily shrink with louder signals.
In at least some cases, more sensitive measurements probe later times without necessarily providing tighter constraints on ringdown frequencies and damping times.
Preliminary investigations are unable to explain this result in terms of a numerical relativity artifact.
\end{abstract}

\maketitle

\section{Introduction}
The no-hair theorem is a remarkable prediction of general relativity (\GR), which states that black holes are described by only three parameters.
For astrophysical black holes there are only two: mass and dimensionless spin.
The resulting spacetime is described by the Kerr metric.
It has long been recognized that gravitational waves may provide an opportunity to test the no-hair theorem; see, e.g.,~\cite{dreyer}.
The basic idea is that the remnant black hole created following a merger event rings down with characteristic frequencies and damping times determined entirely by the mass and spin of the black hole.
By testing that post-merger black holes ring at the correct frequencies and damping times, it ought to be possible to test the validity of the no-hair theorem.

The recent detections of gravitational waves from stellar-mass black hole mergers~\cite{abbott16_detection,boxing_day,gw170104,gw170814} would seem to suggest that a test of the no-hair theorem might be around the corner.
Observational papers already place constraints on parameters of black hole ringdowns~\cite{abbott16_testGR,abbott16_O1BBH,ghosh}.
A number of recent papers highlight the possibilities of ringdown measurements afforded by the expected treasure trove of upcoming merger detections~\cite{bhagwat16,Maselli,yang,Sakai}.
This recent work builds on an already significant body of research on tests of the no-hair theorem, e.g.,~\cite{berti06,Meidam14,gossan}.

Perturbations of the Kerr metric result in gravitational waves given by a sum of damped sinusoids:
\begin{equation}\label{eq:Kerr}
	h(t) = \sum_{\ell m} 
    c_{\ell m} e^{-t/\tau_{\ell m}}
    \sin(2\pi f_{\ell m}t+ \phi_{\ell m}) .
\end{equation}
The sum runs over spheroidal harmonic mode parameters $\ell m$.
The dominant mode is $\ell m=22$.
The next-leading order mode depends on details of the astrophysical system, but, for a merger event, it can be $\ell m=33$.
Both $c_{\ell m}$ and $\phi_{\ell m}$ depend on the details of how the black hole is perturbed.
In contrast, the ringdown frequencies $f_{\ell m}$ and damping times $\tau_{\ell m}$ depend only on the mass and spin of the remnant black hole.
In this way, the no-hair theorem places stringent requirements on the asymptotic behavior of perturbed black holes.
In this paper, the part of the gravitational-wave signal described by~Eq.~\ref{eq:Kerr} is said to be associated with the ``perturbative state."

We note that Eq.~\ref{eq:Kerr} employs a simplifying assumption.
In addition to the sum over $\ell m$ modes, black hole perturbation theory requires for additional sum over $n$ tones.
Equation ~\ref{eq:Kerr} assumes that the signal consists of the $n=0$ primary tones and that the $n\geq1$ overtones are negligible.
This is a reasonable assumption because the damping times of the overtones are typically small compared to the primary tones and so Eq.~\ref{eq:Kerr} becomes an accurate description at late times.
For the remnant of GW150914~\cite{abbott16_detection}, for example, \GR\ predicts primary tones~\cite{berti06} of $(\tau_{220},\tau_{330})=(\unit[3.6]{ms},\unit[3.5]{ms})$ versus overtones of $(\tau_{221},\tau_{331})=(\unit[1.2]{ms},\unit[1.2]{ms})$.

The above assumption is not only reasonable, it is also necessary for our present task.
Numerical relativity simulations are able to isolate different $\ell m$ modes, but they do not provide a means of separating out the contributions from different tones.
This means that we have no way of telling the difference between the presence of superpositions of {\em linear} overtones and {\em non-linear} perturbations leftover from the merger.
Since the no-hair theorem concerns itself with linear perturbations, the safe course of action seems to be to focus exclusively on the primary tones.

While the post-merger remnant approaches the perturbative state asymptotically, at no point in time is the post-merger waveform precisely described by the perturbative state.
At finite times, there is always a deviation, however small, left over from the merger.
Moreover, the ringdown signal becomes weaker as it settles into the form described by Eq.~\ref{eq:Kerr}.
This leads to tension: on the one hand we want to maximize the signal-to-noise ratio of our observation.
On the other hand, we want to wait for the remnant black hole to settle to the perturbative state where the no-hair theorem applies.
In this paper, we show that this tension leads to surprising scaling relations.
For some merger events, confidence intervals on ringdown parameters do not necessarily shrink monotonically with increasing loudness.

The remainder of this paper is organized as follows.
First, we introduce the \GRP\ formalism, designed to isolate the part of black hole ringdown waveform that is indistinguishable from the perturbative regime.
This establishes a framework in which we can carry out unbiased parameter estimation.
We investigate how constraints on ringdown parameters scale with the loudness of the signal.
We document how confidence intervals on $f_{\ell m}, \tau_{\ell m}$ scale with increasingly loud signals.
We conclude with a discussion of the implications.

\section{Parameter estimation with \GRP}
Following a binary merger, the frequency and damping time of a black hole ringdown are entirely determined by the properties of the progenitor binary.
That is, $(f_{\ell m},\tau_{\ell m})$ are not free parameters.
In order to be able to treat them as such, it is necessary to introduce a parameterization.
The \GRP\ parameterization states that each $\ell m$ mode can be written as
\begin{align}\label{eq:GR+}
	h_{\ell m}^\text{GR+}&(t; f_{\ell m}, \tau_{\ell m}) =\notag\\
    &\left\{
    \begin{array}{ll}
     h_{\ell m}^\text{GR}(t) & t<t^\text{cut}_{\ell m} \\
     c_{\ell m} e^{-t/\tau_{\ell m}}
     \sin(2\pi f_{\ell m} t + \phi_{\ell m}) & t>t^\text{cut}_{\ell m} 
    \end{array}
    \right..
\end{align}
The first part of the waveform, up to time $t^\text{cut}_{\ell m}$, is described by $h_{\ell m}^\text{GR}(t)$---the waveform predicted by \GR\ (and calculated with numerical relativity).
After $t^\text{cut}_{\ell m}$, the waveform is described by a damped sine.
The amplitude $c_{\ell m}$ and phase $\phi_{\ell m}$ are determined by requiring continuity of $h_{\ell m}^\text{GR+}(t)$ and its first derivative at $t^\text{cut}_{\ell m}$.

We determine each $t^\text{cut}_{\ell m}$ by insisting that the parameterized component of the waveform is applied only after the remnant black hole has settled into the perturbative state---as ascertained by measurement with a gravitational-wave detector.
In particular, we require that $h^\text{GR+}_{\ell m}(t; f^\text{GR}_{\ell m},\tau^\text{GR}_{\ell m})$ is indistinguishable from $h^\text{GR}_{\ell m}(t)$. 
In the frequentist framework, the time series $h^\text{GR+}_{\ell m}$ and $h^\text{GR}_{\ell m}$ are indistinguishable if the residuals 
\begin{equation}
	\delta h_{\ell m} \equiv h^\text{GR+}_{\ell m}-h^\text{GR}_{\ell m} ,
\end{equation}
are not detectable.
We can use a matched filter template to detect non-zero residuals $\delta h_{\ell m}$ in our data.
The expectation value of the signal-to-noise ratio for our matched filter is given by
\begin{equation}\label{eq:D}
	D_{\ell m} \equiv (\delta h_{\ell m}| \delta h_{\ell m})^{1/2} .
\end{equation}
The parentheses denote an inner product
\begin{equation}
  \left(a|b\right)\equiv4\text{Re}\sum_{k=1}^M \frac{\tilde{a}(f)\tilde{b}^\star(f)}{\sigma_h^2(f)} ,
\end{equation}
where $\sigma_h(f)$ is the noise amplitude spectral density.
The index $k$ labels frequency bins of which there are $M$.
In order to define the perturbative portion of the waveform, we require: $D_{\ell m}< 1$.

Note that the value of $t^\text{cut}_{\ell m}$ is {\em detector dependent}.
The more sensitive the measurement,  the larger the value of $t^\text{cut}_{\ell m}$.
There is a different value of $t_{\ell m}^\text{cut}$ for each $\ell m$ mode, the set of which form a vector denoted $\pmb{t}^\text{cut}$.
Similarly, we introduce vectors $\pmb{f}$ and $\pmb\tau$.
The \GR\ waveform is practically indistinguishable from \GRP\ evaluated at $\pmb{f}^\text{GR}, \pmb\tau^\text{GR}$.
This method of choosing $\pmb{t}^\text{cut}$ ought to produce the smallest possible $(\pmb{f},\pmb\tau)$ confidence intervals subject to the constraint that any bias---arising from the fact that the \GR\ waveform is not a perfect exponentially damped sinusoid---is small.
Some of the ingredients for \GRP\ are already in the literature.
For example, previous observational results on the ringdown parameters $(f_{22}, \tau_{22})$ provide constraints as a function of $t_{22}^\text{cut}$; see, e.g., Fig.~5 of~\cite{abbott16_testGR}.
One could choose the confidence interval in this figure corresponding to the \GRP\ value of $t^\text{cut}_{22}$. 

One may ask if Eq.~\ref{eq:D} provides a suitable method for determining $t^\text{cut}_{\ell m}$.
If we were to choose $t^\text{cut}_{\ell m}$ such that $D_{\ell m}\gg1$---and assuming that \GR\ is correct---we would see with high statistical confidence that the data are not consistent with the perturbative state described by Eq.~\ref{eq:Kerr}.
It seems undesirable to carry out fits for ringdown parameters using data, which is manifestly inconsistent with a Kerr ringdown.
Therefore, we argue that the $D<1$ method is suitable.

Throughout this paper, we use GW150914 as our fiducial merger event.
We use waveforms~\cite{Lovelace:2016uwp} from the SXS collaboration~\footnote{\url{https://www.black-holes.org/waveforms/}} consistent with the best-fit parameters~\footnote{The precise waveform ID is SXS:BBH:0305, corresponding to mass ratio $m_1/m_2=1.22$, dimensionless spin magnitudes at the relaxation time $\vec\chi_1=(3.4\times10^{-8},-4.1\times10^{-8},0.33	),\vec\chi_2=(3.8\times10^{-8},3.2\times10^{-8},-0.44)$, eccentricity $\epsilon=8.4\times10^{-4}$, and orbital frequency multiplied by the total Christodoulou mass at the relaxation time $M\omega=0.018$. We assume a source observed at $\text{GPS}=969379706$ and $(\textsc{RA},\textsc{DEC})=(\unit[9]{hr},5^\circ)$.} of GW150914~\cite{abbott16_PE}.
For a GW150914-like event at $d=\unit[410]{Mpc}$, and assuming the two-detector LIGO network operating at design sensitivity~\cite{ligo15}, $t^\text{cut}_{22}\approx\unit[8]{ms}$ and $t_{33}^\text{cut}\approx\unit[5]{ms}$.
Of course, if GW150914 had been closer, the signal would have been louder, and these values of $t_{\ell m}^\text{cut}$ would have to be bigger to ensure $D_{\ell m}<1$.

In order to investigate scaling behavior, we introduce a ``loudness'' parameter: 
\begin{equation}\label{eq:loudness}
	\text{loudness} = \frac{\unit[410]{Mpc}}{\text{distance}} \propto \rho \propto \sigma_h^{-1} \propto N^{1/2} .
\end{equation}
Doubling the loudness of the waveform is equivalent to halving the distance $d$ to the event, or alternatively, doubling the matched filter signal-to-noise ratio $\rho$, or equivalently, halving the detector noise $\sigma_h$.
We can also think of boosting the signal by stacking data from an ensemble of GW150914-like mergers.
In this case, loudness scales like the square root of the number of events $N$; see, e.g.,~\cite{yang}.

Using \GRP, and assuming Gaussian noise, the log likelihood function is
\begin{equation}\label{eq:lnl}
	\ln{\cal L}(s|\pmb{f}, {\pmb{\tau}}) \propto
    -\frac{1}{2}\sum_{k=1}^M \frac{|\tilde{s}_k-\tilde{u}_k(\pmb{f},\pmb\tau)|^2}{\sigma_{h,k}^2} ,
\end{equation}
where $\tilde{s}_k=\tilde{h}_k+\tilde{n}_k$ is the (Fourier transform of the) strain data consisting of signal $\tilde{h}_k$ and noise $\tilde{n}_k$.
The variable $\tilde{u}_k(\pmb{f},\pmb\tau)$ is the predicted \GRP\ waveform.
The posteriors for the Kerr ringdown parameters are
\begin{align}\label{eq:f}
	p(\pmb{f} | s) \propto &
    \int d\pmb\tau
    {\cal L}(s|\pmb{f}, {\pmb{\tau}}, \pmb{t}^\text{cut})
    p(\pmb{\tau})
    p(\pmb{f}) \\\label{eq:tau}
    p(\pmb\tau | s) \propto &
    \int d\pmb{f}
    {\cal L}(s|\pmb{f}, {\pmb{\tau}}, \pmb{t}^\text{cut})
    p(\pmb{\tau})
    p(\pmb{f})
\end{align}
Here $p(\pmb\tau)$ and $p(\pmb{f})$ are priors on ringdown parameters, which we take to be flat~\footnote{For the sake of simplicity, we have ignored uncertainty in other astrophysical parameters, which introduces uncertainty into the matching parameters $c_{\ell m}, \phi_{\ell m}$; see Eq.~\ref{eq:GR+}.
In a more careful treatment, it is necessary to marginalize over these additional parameters.
Since we do not include this uncertainty, our confidence intervals on $(\pmb{f},\pmb\tau)$ are optimistic.}.
We calculate posteriors using MultiNest~\cite{MultiNest,PyMultiNest}.

\section{Results}
We are now ready to use the \GRP\ formalism to see how constraints on $(\pmb{f},\pmb\tau)$ scale with loudness.
The results are pertinent if we wish to know what kind of measurement is required in order for detectors such as LIGO to measure ringdown parameters to some tolerance, thereby validating the no-hair theorem; see, e.g.,~\cite{Sakai}.
We consider a range of loudness between $(1,45)$.
For each value of loudness, we determine $t^\text{cut}_{\ell m}$ using Eq.~\ref{eq:D}.
We perform separate calculations for the $\ell m=22$ and $\ell m=33$ modes.
For the sake of simplicity, we ignore the difficulties that arise from trying to separate these two modes and assume they can be isolated; see~\cite{bhagwat16}.
Using $t^\text{cut}_{\ell m}$, we calculate posteriors for $(f_{\ell m}, \tau_{\ell m})$ with Eqs.~\ref{eq:f}-\ref{eq:tau}, which, in turn, we use to derive 95\% confidence intervals.
The confidence intervals are calculated assuming Gaussian, Advanced LIGO design-sensitivity noise~\cite{ligo15}.
Following common practice, we use the median noise realization such that $\tilde{n}_k=0$.

\begin{figure*}
    \begin{minipage}{0.49\textwidth}
	\includegraphics[width=\textwidth]{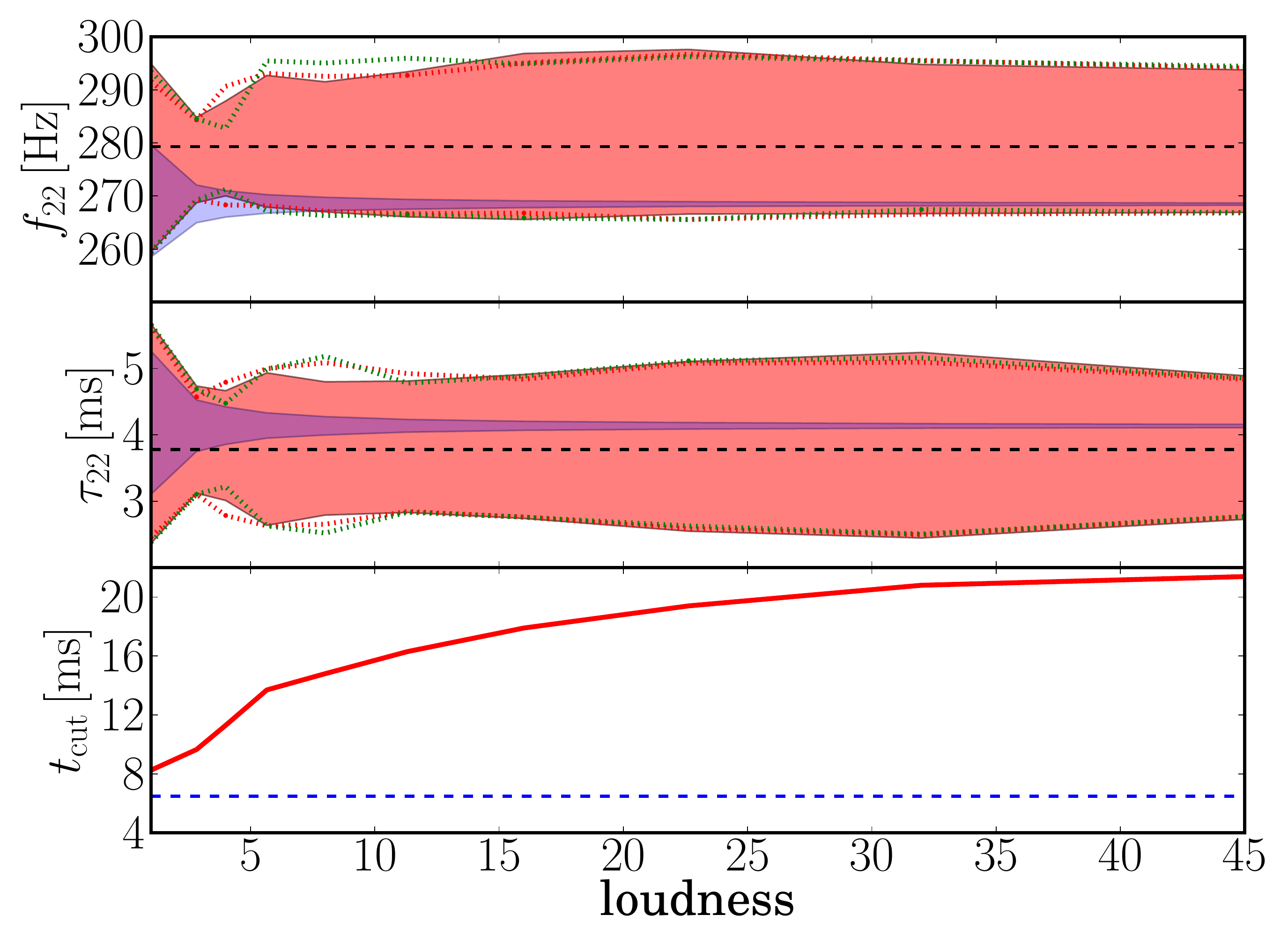}
    \end{minipage}
    \begin{minipage}{0.49\textwidth}
    \includegraphics[width=\textwidth]{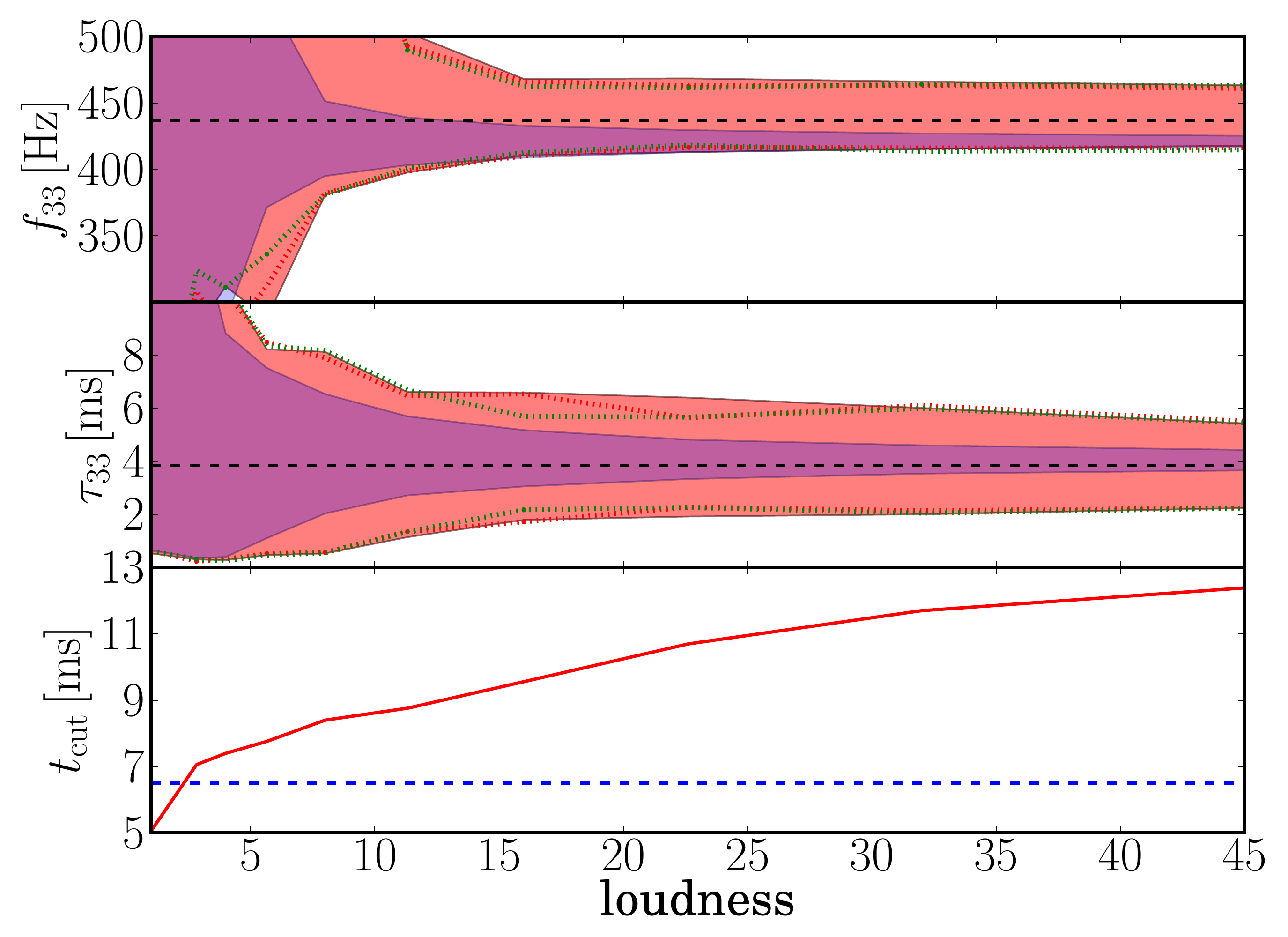}
    \end{minipage}
	\caption{
    Scaling of \GRP\ confidence intervals with loudness (defined in Eq.~\ref{eq:loudness}).
    The left-hand side are results for $\ell m=22$ while the right-hand side is for $\ell m=33$.
   The top (middle) panels plot ringdown frequency (damping time) as a function of loudness.
   The red shading shows the 95\% confidence regions for \GRP.
   The blue shading shows the 95\% confidence region when $t^\text{cut}_{\ell m}=\unit[6.5]{ms}$.
   The dashed black lines indicate the true parameter value.
	The red \GRP\ interval does not shrink monotonically suggesting a limit to our ability to measure ringdown parameters.
    The blue $t^\text{cut}_{\ell m}=\unit[6.5]{ms}$ interval shrinks monotonically, but exhibits a bias so that the confidence interval eventually excludes the true value.
    The dashed red curves show  the confidence intervals using the lower-resolution L5 waveform.
    The dashed green curves show the confidence intervals taking into account spheroidal-harmonic corrections.
   The bottom panels show $t^\text{cut}_{\ell m}$ (Eq.~\ref{eq:D}).
   }
   \label{fig:loudness}
\end{figure*}

The results are summarized in Fig.~\ref{fig:loudness}.
The left-hand side shows the results for $\ell m=22$ while the right shows results for $\ell m=33$.
Each panel is a function of loudness (Eq.~\ref{eq:loudness}).
The top (middle) panel shows the 95\% confidence intervals on $f_{\ell m}$ ($\tau_{\ell m}$) obtained with \GRP\ in red.
The true value of $f_{\ell m}$ ($\tau_{\ell m}$) is indicated with dashed black lines.
The \GRP\ confidence interval does not shrink monotonically with loudness.

The shaded blue region shows the confidence interval we obtain by arbitrarily setting $t^\text{cut}_{\ell m}=\unit[6.5]{ms}$.
In contrast to the red \GRP\ confidence interval, the blue $t^\text{cut}_{\ell m}=\unit[6.5]{ms}$ confidence interval shrinks monotonically with loudness.
However, it eventually excludes the true value of $(f_{\ell m},\tau_{\ell m})$ in dashed black because the non-linear part of the waveform gives us a bias estimate of $(f_{\ell m}, \tau_{\ell m})$.
This serves as a reminder of the motivation for introducing \GRP\ in the first place: we require a means of carrying out unbiased parameter estimation.
The bottom panel shows $t^\text{cut}_{\ell m}$.

As noted above, Fig.~\ref{fig:loudness} is generated using numerical relativity waveforms from the SXS collaboration.
The shaded red region, in particular, is calculated using the highest-resolution L6 waveform.
In order to test if the non-monotonic scaling is a result of a numerical relativity artifact, we repeat the calculation using the lower-resolution L5 waveform (dashed red line).
The L5 waveform employs a spectral adaptive-mesh-refinement error tolerance that is a factor of $e$ larger than that of L6.
The dashed red curves track the solid red.
The most significant disagreement, near $\text{loudness}=8$ is slight.
If the scaling were the result of a numerical relativity artifact, we would have expected a more significant change.

\section{Discussion}
Can the scaling behavior in Fig.~\ref{fig:loudness} be explained in terms of a numerical relativity effect?
We comment on a few possibilities.
First, Boyle has pointed out that small drifts in the center-of-mass coordinate lead to Bondi-Metzner-Sachs (BMS) supertranslations, which induce mode mixing~\cite{Boyle}.
For the waveform considered here, this effect is estimated to be negligible, though, perhaps small, uncorrected center-of-mass motion is sufficiently large to produce the scaling observed in Fig.~\ref{fig:loudness}.

\begin{figure}
 \includegraphics[width=0.5\textwidth]{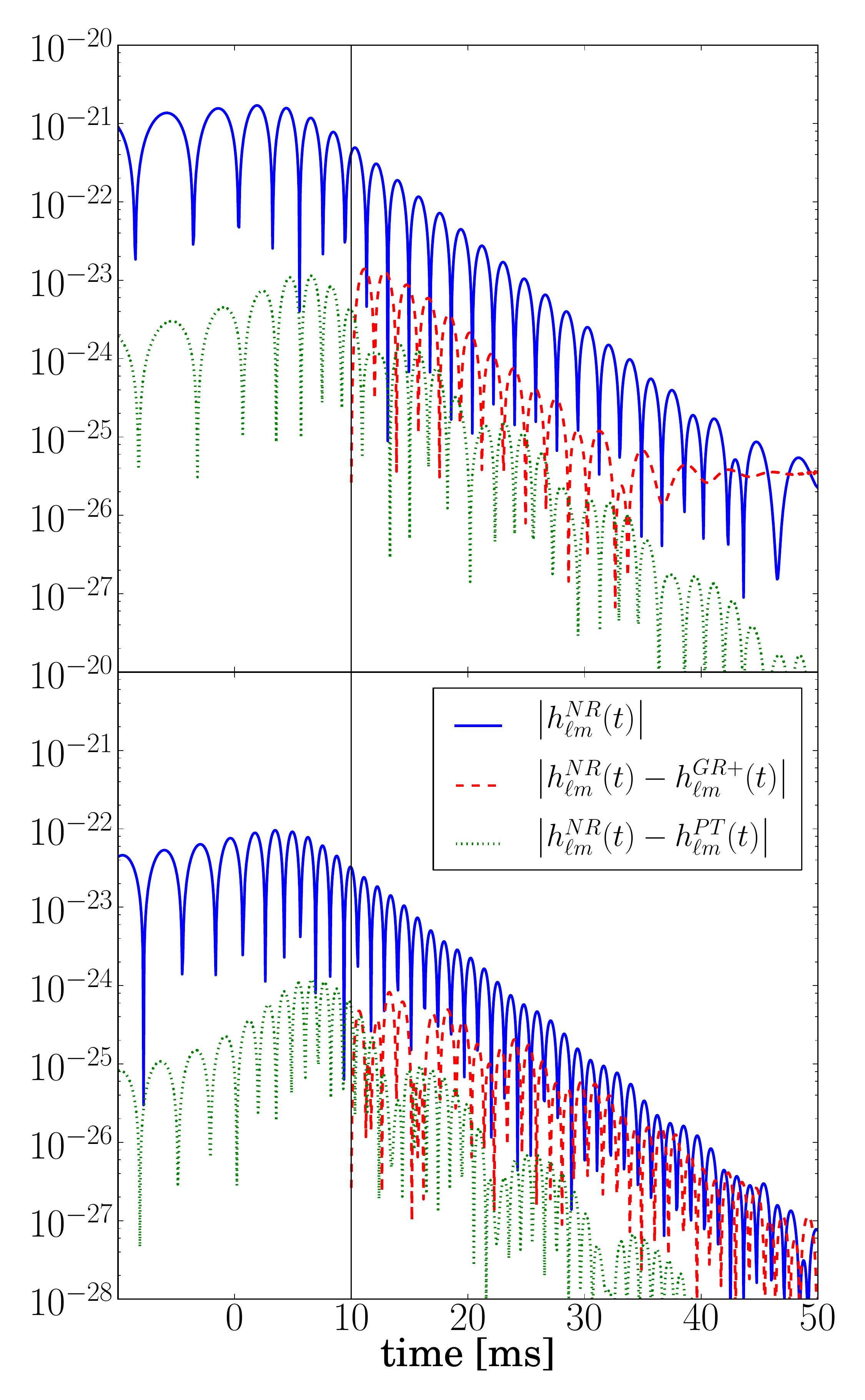}
	\caption{
	The absolute value of the strain time series for the $\ell m = 22$ mode (top) and $\ell m = 33$ mode (bottom).
    The solid blue curves show the  numerical relativity waveforms $|h^\text{NR}_{\ell m}(t)|$ from~\cite{Lovelace:2016uwp}.
    The dashed red curves show the residuals of the \GRP\ fit $|h^\text{NR}_{\ell m}(t)-h^\text{GR+}_{\ell m}(t)|$ with $t^\text{cut}=\unit[10]{ms}$.
    The dotted green curves show the residuals of the spheroidal-spherical mismatch $|h^\text{NR}_{\ell m}(t)-h^\text{PT}_{\ell m}(t)|$.
	Deviation from linear perturbation theory is visible by eye in the envelope of the $33$ mode (bottom blue).
 }
   \label{fig:plot_log_ht}
\end{figure}

Second, each ringdown mode is associated with a different spin-weighted spheroidal harmonic ${}_{-2}S_{\ell m}$.
Numerical relativity extraction, however, is typically carried out with {\em spherical} harmonics ${}_{-2}Y_{\ell m}$.
London et al.\ have pointed out that conflating spheroidal and spherical harmonics can lead to non-negligible mode-mixing~\cite{London}.
One can show that pure, perturbation-theory  modes $h^\text{PT}_{\ell m}$ are linear combinations of numerical relativity waveforms $h^\text{NR}_{\ell m}$ extracted with spherical harmonics~\cite{PressTeukolsky}.
Since
\begin{equation}
	h = h_+ + ih_\times = 
    \sum_{\ell m} h_{\ell m}^\text{NR} {}_{-2}Y_{\ell m} = 
    \sum_{\ell' m} h_{\ell m}^\text{PT} {}_{-2}S_{\ell' m} ,
\end{equation}
it follows from the orthogonality of ${}_{-2}Y_{\ell m}$ that
\begin{equation}
	h^\text{NR}_{\ell m} =
    \sum_{\ell'} h_{\ell' m}^\text{PT} \,
    \kappa_{\ell m \ell'} , 
\end{equation}
where
$
	\kappa_{\ell m \ell'} \equiv
	\int d\Omega\,
	{}_{-2}Y_{\ell m}^* {}_{-2}S_{\ell' m}
$
is an integral over solid angle $\Omega$.
We can therefore write, e.g.,
\begin{align}
h_{22}^\text{NR} = & 
\kappa_{222} \, h_{22}^\text{PT} +
\kappa_{223} \, h_{32}^\text{PT} + 
\kappa_{224} \, h_{42}^\text{PT} + ...
\end{align} 
In this expression, $\kappa_{222}$ is close to unity and the other $\kappa_{\ell m \ell'}$ are small.
Working to first order in small $\kappa_{\ell m \ell'}$, we can make the approximations $h_{32}^\text{PT}\approx h_{32}^\text{NR}$ and $h_{42}^\text{PT}\approx h_{42}^\text{NR}$.
Thus,
\begin{align}
h_{22}^\text{PT} \approx & 
\kappa_{222}^{-1} \, h_{22}^\text{NR} -
\kappa_{223}/\kappa_{222} \, h_{32}^\text{NR} - 
\kappa_{224}/\kappa_{222} \, h_{42}^\text{NR} \\
h_{33}^\text{PT} \approx & 
\kappa_{333}^{-1} \, h_{33}^\text{NR} -
\kappa_{334}/\kappa_{333} \, h_{43}^\text{NR} - 
\kappa_{335}/\kappa_{333} \, h_{53}^\text{NR} ,
\end{align} 
In order to estimate the size of this effect, we calculate~\footnote{Using the \textsc{LAL} code \textsc{lalsim-bh-sphwf}, we obtain $\kappa_{222}= 0.99 - 0.03i$, $\kappa_{223}= -0.11 + 0.01i$, $\kappa_{224}= 0.020 - 0.02i$, $\kappa_{333}=1.00 - 0.02i$, $\kappa_{334}=-0.11 + 0.008i$, and $\kappa_{335}=0.0034 + 0.0006i$.} $\kappa_{\ell m \ell'}$ using publicly available code in the LSC Algorithm Library (\textsc{LAL})~\footnote{\url{https://wiki.ligo.org/DASWG/LALSuite}}.
We repeat the analysis using $h_{\ell m}^\text{PT}$.
The dashed green curve in Fig.~\ref{fig:loudness} shows the resulting confidence intervals.
We observe only a marginal change, suggesting that spheroidal-spherical mismatch is not responsible for the scaling behavior.
The relative smallness of this effect is highlighted in Fig.~\ref{fig:plot_log_ht} where we compare $h^\text{PT}_{\ell m}(t)$ with $h^\text{NR}_{\ell m}(t)$.

Third, there are a number of other systematic effects, which could in principle complicate the characterization of ringdown parameters including non-linear memory~\cite{Christodoulou} and late-time power-law tails~\cite{Price}.
Neither of these seem to us to be likely explanations for Fig.~\ref{fig:loudness}.
They seem too small and the time scales do not fit.
We also attempted to account for ringdown back-reaction, in which the mass and spin of the black hole change over time due to the emission of gravitational waves.
Waveforms designed to take into account ringdown back-reaction did not yield superior fits.

\section{Conclusions}
Motivated by the desire to rigorously test the no-hair theorem in the domain in which it applies, we introduce the \GRP. This formalism enables us to carry out unbiased parameter estimation of black hole ringdowns.
The \GRP\ formalism provides a method for testing the no-hair theorem using only data from after the remnant black hole has settled into a perturbative state.

We investigate how \GRP\ confidence regions scale with signal loudness and observe non-monotonic behavior.
By insisting that the remnant black hole settles into the perturbative state, louder signals can, in at least some cases, probe later times without necessarily yielding tighter constraints on ringdown parameters.
It is not clear the extent to which this behavior can be attributed to a numerical relativity artifact, in which case it might be possible to remedy.
A less appealing alternative hypothesis is that residual non-linearity in the post-merger signal decays on a timescale comparable to the linear signal we seek to measure.

It is hard to say conclusively that this effect is physical and not a numerical relativity artifact.
However, since we observe comparable scaling behavior in two waveforms from simulations with different grid resolutions, we currently have no evidence in favor of the numerical relativity error hypothesis.
Further work should be carried out to see if this scaling holds for additional numerical relativity waveforms calculated using different prescriptions and with higher resolution.

We thank Jolien Creighton, Colin Capano, Alessandra Buonanno, Vivien Raymond, and Nathan Johnson-McDaniel for helpful discussion.
ET is supported through ARC FT150100281 and CE170100004.
PDL is supported through ARC FT160100112.
YL is supported through ARC CE170100004.
This is document LIGO-P1700136.

\bibliography{TestingKerr}

\begin{thebibliography}{25}
\expandafter\ifx\csname natexlab\endcsname\relax\def\natexlab#1{#1}\fi
\expandafter\ifx\csname bibnamefont\endcsname\relax
  \def\bibnamefont#1{#1}\fi
\expandafter\ifx\csname bibfnamefont\endcsname\relax
  \def\bibfnamefont#1{#1}\fi
\expandafter\ifx\csname citenamefont\endcsname\relax
  \def\citenamefont#1{#1}\fi
\expandafter\ifx\csname url\endcsname\relax
  \def\url#1{\texttt{#1}}\fi
\expandafter\ifx\csname urlprefix\endcsname\relax\def\urlprefix{URL }\fi
\providecommand{\bibinfo}[2]{#2}
\providecommand{\eprint}[2][]{\url{#2}}

\bibitem[{\citenamefont{Dreyer et~al.}(2004)\citenamefont{Dreyer, Kelly,
  Krishnan, Finn, Garrison, and Lopez-Aleman}}]{dreyer}
\bibinfo{author}{\bibfnamefont{O.}~\bibnamefont{Dreyer}},
  \bibinfo{author}{\bibfnamefont{B.}~\bibnamefont{Kelly}},
  \bibinfo{author}{\bibfnamefont{B.}~\bibnamefont{Krishnan}},
  \bibinfo{author}{\bibfnamefont{L.~S.} \bibnamefont{Finn}},
  \bibinfo{author}{\bibfnamefont{D.}~\bibnamefont{Garrison}}, \bibnamefont{and}
  \bibinfo{author}{\bibfnamefont{R.}~\bibnamefont{Lopez-Aleman}},
  \bibinfo{journal}{Class. Quant. Grav.} \textbf{\bibinfo{volume}{21}},
  \bibinfo{pages}{787} (\bibinfo{year}{2004}).

\bibitem[{\citenamefont{{Abbott} et~al.}(2016)}]{abbott16_detection}
\bibinfo{author}{\bibfnamefont{B.~P.} \bibnamefont{{Abbott}}}
  \bibnamefont{et~al.}, \bibinfo{journal}{Phys. Rev. Lett.}
  \textbf{\bibinfo{volume}{116}}, \bibinfo{pages}{061102}
  (\bibinfo{year}{2016}).

\bibitem[{\citenamefont{Abbott et~al.}(2016)}]{boxing_day}
\bibinfo{author}{\bibfnamefont{B.~P.} \bibnamefont{Abbott}}
  \bibnamefont{et~al.}, \bibinfo{journal}{Phys. Rev. Lett.}
  \textbf{\bibinfo{volume}{116}}, \bibinfo{pages}{241103}
  (\bibinfo{year}{2016}).

\bibitem[{\citenamefont{Abbott et~al.}(2017{\natexlab{a}})}]{gw170104}
\bibinfo{author}{\bibfnamefont{B.~P.} \bibnamefont{Abbott}}
  \bibnamefont{et~al.}, \bibinfo{journal}{Phys. Rev. Lett.}
  \textbf{\bibinfo{volume}{118}}, \bibinfo{pages}{221101}
  (\bibinfo{year}{2017}{\natexlab{a}}).

\bibitem[{\citenamefont{Abbott et~al.}(2017{\natexlab{b}})}]{gw170814}
\bibinfo{author}{\bibfnamefont{B.~P.} \bibnamefont{Abbott}}
  \bibnamefont{et~al.}, \bibinfo{journal}{Phys. Rev. Lett.}
  \textbf{\bibinfo{volume}{119}}, \bibinfo{pages}{141101}
  (\bibinfo{year}{2017}{\natexlab{b}}).

\bibitem[{\citenamefont{{Abbott} et~al.}(2016{\natexlab{a}})}]{abbott16_testGR}
\bibinfo{author}{\bibfnamefont{B.~P.} \bibnamefont{{Abbott}}}
  \bibnamefont{et~al.}, \bibinfo{journal}{Phys. Rev. Lett.}
  \textbf{\bibinfo{volume}{116}}, \bibinfo{pages}{221101}
  (\bibinfo{year}{2016}{\natexlab{a}}).

\bibitem[{\citenamefont{{Abbott} et~al.}(2016{\natexlab{b}})}]{abbott16_O1BBH}
\bibinfo{author}{\bibfnamefont{B.~P.} \bibnamefont{{Abbott}}}
  \bibnamefont{et~al.}, \textbf{\bibinfo{volume}{6}}, \bibinfo{pages}{041015}
  (\bibinfo{year}{2016}{\natexlab{b}}).

\bibitem[{\citenamefont{Ghosh et~al.}(2016)\citenamefont{Ghosh, Ghosh,
  Johnson-McDaniel, Mishra, Ajith, Pozzo, Nichols, Chen, Nielsen, Berry
  et~al.}}]{ghosh}
\bibinfo{author}{\bibfnamefont{A.}~\bibnamefont{Ghosh}},
  \bibinfo{author}{\bibfnamefont{A.}~\bibnamefont{Ghosh}},
  \bibinfo{author}{\bibfnamefont{N.~K.} \bibnamefont{Johnson-McDaniel}},
  \bibinfo{author}{\bibfnamefont{C.~K.} \bibnamefont{Mishra}},
  \bibinfo{author}{\bibfnamefont{P.}~\bibnamefont{Ajith}},
  \bibinfo{author}{\bibfnamefont{W.~D.} \bibnamefont{Pozzo}},
  \bibinfo{author}{\bibfnamefont{D.~A.} \bibnamefont{Nichols}},
  \bibinfo{author}{\bibfnamefont{Y.}~\bibnamefont{Chen}},
  \bibinfo{author}{\bibfnamefont{A.~B.} \bibnamefont{Nielsen}},
  \bibinfo{author}{\bibfnamefont{C.~P.~L.} \bibnamefont{Berry}},
  \bibnamefont{et~al.}, \bibinfo{journal}{Phys. Rev. D}
  \textbf{\bibinfo{volume}{94}}, \bibinfo{pages}{021101}
  (\bibinfo{year}{2016}).

\bibitem[{\citenamefont{Bhagwat et~al.}(2016)\citenamefont{Bhagwat, Brown, and
  Ballmer}}]{bhagwat16}
\bibinfo{author}{\bibfnamefont{S.}~\bibnamefont{Bhagwat}},
  \bibinfo{author}{\bibfnamefont{D.~A.} \bibnamefont{Brown}}, \bibnamefont{and}
  \bibinfo{author}{\bibfnamefont{S.~W.} \bibnamefont{Ballmer}},
  \bibinfo{journal}{Phys. Rev. D} \textbf{\bibinfo{volume}{94}},
  \bibinfo{pages}{084024} (\bibinfo{year}{2016}).

\bibitem[{\citenamefont{Maselli et~al.}(2017)\citenamefont{Maselli, Kokkotas,
  and Laguna}}]{Maselli}
\bibinfo{author}{\bibfnamefont{A.}~\bibnamefont{Maselli}},
  \bibinfo{author}{\bibfnamefont{K.}~\bibnamefont{Kokkotas}}, \bibnamefont{and}
  \bibinfo{author}{\bibfnamefont{P.}~\bibnamefont{Laguna}}
  (\bibinfo{year}{2017}), \bibinfo{note}{https://arxiv.org/abs/1702.01110}.

\bibitem[{\citenamefont{Yang et~al.}(2017)\citenamefont{Yang, Yagi, Blackman,
  Luis~Lehner, Pretorius, and Yunes}}]{yang}
\bibinfo{author}{\bibfnamefont{H.}~\bibnamefont{Yang}},
  \bibinfo{author}{\bibfnamefont{K.}~\bibnamefont{Yagi}},
  \bibinfo{author}{\bibfnamefont{J.}~\bibnamefont{Blackman}},
  \bibinfo{author}{\bibfnamefont{V.~P.} \bibnamefont{Luis~Lehner}},
  \bibinfo{author}{\bibfnamefont{F.}~\bibnamefont{Pretorius}},
  \bibnamefont{and} \bibinfo{author}{\bibfnamefont{N.}~\bibnamefont{Yunes}},
  \bibinfo{journal}{Phys. Rev. Lett.} \textbf{\bibinfo{volume}{118}},
  \bibinfo{pages}{161101} (\bibinfo{year}{2017}).

\bibitem[{\citenamefont{Sakai et~al.}(2017)\citenamefont{Sakai, Oohara, Nakano,
  Kaneyama, and Takahashi}}]{Sakai}
\bibinfo{author}{\bibfnamefont{K.}~\bibnamefont{Sakai}},
  \bibinfo{author}{\bibfnamefont{K.-I.} \bibnamefont{Oohara}},
  \bibinfo{author}{\bibfnamefont{H.}~\bibnamefont{Nakano}},
  \bibinfo{author}{\bibfnamefont{M.}~\bibnamefont{Kaneyama}}, \bibnamefont{and}
  \bibinfo{author}{\bibfnamefont{H.}~\bibnamefont{Takahashi}}
  (\bibinfo{year}{2017}), \bibinfo{note}{https://arxiv.org/abs/1705.04107}.

\bibitem[{\citenamefont{{Berti} et~al.}(2006)\citenamefont{{Berti}, {Cardoso},
  and {Will}}}]{berti06}
\bibinfo{author}{\bibfnamefont{E.}~\bibnamefont{{Berti}}},
  \bibinfo{author}{\bibfnamefont{V.}~\bibnamefont{{Cardoso}}},
  \bibnamefont{and} \bibinfo{author}{\bibfnamefont{C.~M.}
  \bibnamefont{{Will}}}, \bibinfo{journal}{Phys. Rev. D}
  \textbf{\bibinfo{volume}{73}}, \bibinfo{pages}{064030}
  (\bibinfo{year}{2006}).

\bibitem[{\citenamefont{Meidam et~al.}(2014)\citenamefont{Meidam, Agathos, {Van
  Den Broeck}, Veitch, and Sathyaprakash}}]{Meidam14}
\bibinfo{author}{\bibfnamefont{J.}~\bibnamefont{Meidam}},
  \bibinfo{author}{\bibfnamefont{M.}~\bibnamefont{Agathos}},
  \bibinfo{author}{\bibfnamefont{C.}~\bibnamefont{{Van Den Broeck}}},
  \bibinfo{author}{\bibfnamefont{J.}~\bibnamefont{Veitch}}, \bibnamefont{and}
  \bibinfo{author}{\bibfnamefont{B.~S.} \bibnamefont{Sathyaprakash}},
  \bibinfo{journal}{Phys. Rev. D} \textbf{\bibinfo{volume}{90}},
  \bibinfo{pages}{064009} (\bibinfo{year}{2014}).

\bibitem[{\citenamefont{Gossan et~al.}(2012)\citenamefont{Gossan, Veitch, and
  Sathyaprakash}}]{gossan}
\bibinfo{author}{\bibfnamefont{S.}~\bibnamefont{Gossan}},
  \bibinfo{author}{\bibfnamefont{J.}~\bibnamefont{Veitch}}, \bibnamefont{and}
  \bibinfo{author}{\bibfnamefont{B.~S.} \bibnamefont{Sathyaprakash}},
  \bibinfo{journal}{Phys. Rev. D} \textbf{\bibinfo{volume}{85}},
  \bibinfo{pages}{124056} (\bibinfo{year}{2012}).

\bibitem[{\citenamefont{Lovelace et~al.}(2016)}]{Lovelace:2016uwp}
\bibinfo{author}{\bibfnamefont{G.}~\bibnamefont{Lovelace}}
  \bibnamefont{et~al.}, \bibinfo{journal}{Class. Quant. Grav.}
  \textbf{\bibinfo{volume}{33}}, \bibinfo{pages}{244002}
  (\bibinfo{year}{2016}).

\bibitem[{\citenamefont{{Abbott} et~al.}(2016{\natexlab{c}})}]{abbott16_PE}
\bibinfo{author}{\bibfnamefont{B.~P.} \bibnamefont{{Abbott}}}
  \bibnamefont{et~al.}, \bibinfo{journal}{Phys. Rev. Lett.}
  \textbf{\bibinfo{volume}{116}}, \bibinfo{pages}{241102}
  (\bibinfo{year}{2016}{\natexlab{c}}).

\bibitem[{\citenamefont{Aasi et~al.}(2015)}]{ligo15}
\bibinfo{author}{\bibfnamefont{J.}~\bibnamefont{Aasi}} \bibnamefont{et~al.},
  \bibinfo{journal}{Classical and Quantum Gravity}
  \textbf{\bibinfo{volume}{32}}, \bibinfo{pages}{074001}
  (\bibinfo{year}{2015}).

\bibitem[{\citenamefont{Feroz et~al.}(2009)\citenamefont{Feroz, Hobson, and
  Bridges}}]{MultiNest}
\bibinfo{author}{\bibfnamefont{F.}~\bibnamefont{Feroz}},
  \bibinfo{author}{\bibfnamefont{M.~P.} \bibnamefont{Hobson}},
  \bibnamefont{and} \bibinfo{author}{\bibfnamefont{M.}~\bibnamefont{Bridges}},
  \bibinfo{journal}{MNRAS} \textbf{\bibinfo{volume}{398}},
  \bibinfo{pages}{1601} (\bibinfo{year}{2009}).

\bibitem[{\citenamefont{Buchner et~al.}(2014)}]{PyMultiNest}
\bibinfo{author}{\bibfnamefont{J.}~\bibnamefont{Buchner}} \bibnamefont{et~al.},
  \bibinfo{journal}{A\&A} \textbf{\bibinfo{volume}{564}}, \bibinfo{pages}{A125}
  (\bibinfo{year}{2014}).

\bibitem[{\citenamefont{Boyle}(2016)}]{Boyle}
\bibinfo{author}{\bibfnamefont{M.}~\bibnamefont{Boyle}},
  \bibinfo{journal}{Phys. Rev. D} \textbf{\bibinfo{volume}{93}},
  \bibinfo{pages}{084031} (\bibinfo{year}{2016}).

\bibitem[{\citenamefont{London et~al.}(2014)\citenamefont{London, Shoemaker,
  and Healy}}]{London}
\bibinfo{author}{\bibfnamefont{L.}~\bibnamefont{London}},
  \bibinfo{author}{\bibfnamefont{D.}~\bibnamefont{Shoemaker}},
  \bibnamefont{and} \bibinfo{author}{\bibfnamefont{J.}~\bibnamefont{Healy}},
  \bibinfo{journal}{Phys. Rev. D} \textbf{\bibinfo{volume}{90}},
  \bibinfo{pages}{124032} (\bibinfo{year}{2014}).

\bibitem[{\citenamefont{Press and Teukolsky}(1973)}]{PressTeukolsky}
\bibinfo{author}{\bibfnamefont{W.~H.} \bibnamefont{Press}} \bibnamefont{and}
  \bibinfo{author}{\bibfnamefont{S.~A.} \bibnamefont{Teukolsky}},
  \bibinfo{journal}{Astrophys. J.} \textbf{\bibinfo{volume}{185}},
  \bibinfo{pages}{649} (\bibinfo{year}{1973}).

\bibitem[{\citenamefont{Christodoulou}(1991)}]{Christodoulou}
\bibinfo{author}{\bibfnamefont{D.}~\bibnamefont{Christodoulou}},
  \bibinfo{journal}{Phys. Rev. Lett.} \textbf{\bibinfo{volume}{67}},
  \bibinfo{pages}{1486} (\bibinfo{year}{1991}).

\bibitem[{\citenamefont{Price}(1972)}]{Price}
\bibinfo{author}{\bibfnamefont{R.~H.} \bibnamefont{Price}},
  \bibinfo{journal}{Phys. Rev. D} \textbf{\bibinfo{volume}{5}},
  \bibinfo{pages}{2419} (\bibinfo{year}{1972}).

\end{thebibliography}

\end{document}